# A Semi-Intelligence Algorithm For Charging Electric Vehicles

1st A. Mousaei
*Dept. Electrical and Computer Engineering*
*University of Tabriz*
Tabriz, Iran
a.mousaei@tabrizu.ac.ir

2nd H. Allahyari
*Dept. Electrical and Computer Engineering*
*University of KNTU*
Tehran, Iran
h.allahyari@email.kntu.ac.ir

*Abstract*— The penetration of Electric Vehicles (EVs) in Distribution Networks (DNs) has many challenges for distribution companies, such as voltage drop, networklosses, etc. To overcome these problems, it is recommended to use coordinated charging methods. However, thesemethods require high-cost telecommunication, measurement, and processing infrastructure and can only be implemented in smart grids. This article introduces a Semi-Smart Method (SSM) to charge EVs that does not require complex infrastructure. Thismethod, using a simple and inexpensive local automation system, charges EVs during off-peak hours of the DN, thus improving its parameters. Since EVs are charged at low-traffic time intervals, the proposed method benefits EV owners. To get real results, 4-wire DN is considered to model the effect of the neutral wire. To confirm the effectiveness of the proposed method, it is compared to various uncontrolled loading methods. A standard 19-bus test system is used for simulations.

I. INTRODUCTION

The reduction of fossil fuel resources and the expansion of environmental pollutants have caused the expansion of the use of Electric Vehicles (EVs) in the transportation industry. In general, EVs include various structures that Plug-in Hybrid Electric Vehicles (PHEVs) are the most efficient. PHEVs use the Internal Combustion (ICE) system and electric battery in a combined way and have the ability to charge/discharge from/to the electricity Distribution Network (DN).

Despite this, the expansion of more and more PHEVs and their need for electric power has created many problems for electric distribution companies, such as severe voltage fluctuations, increasing losses, and increasing the possibility of blackouts due to overload. This issue is considered an important challenge in terms of demand-side management for electricity distribution companies.

These problems can be solved by renovating the structure of the power systems and creating smart electricity DN and expanding telecommunication and control infrastructures. In fact, in smart grids, EVs are considered controllable loads, whose charging schedules can be programmed. This type of charging of EVs, which is done using infrastructures of communication, measurement, and processing, is called coordinated, controlled, or intelligent charging.

A lot of research has been done on the effect of intelligent charging of EVs on various parameters of the network, such as the voltage profile, losses, reliability, harmonics, etc., and its advantages compared to uncontrolled or random charging have been mentioned. In [9], a model for the controlled charging of EVs is presented in which a three-objective function is used with the goals of minimizing losses, optimizing the load factor, and minimizing load changes. In [10], binary linear programming is used for the problem of synchronous charging of EVs. The reference [11] uses the valley filling strategy for the controlled charging of EVs. The reference [12] for the coordinated charging of EVs, there is a compromise between the total production cost correction. To solve the optimization problem, it uses a decentralized approach. The reference [13] deals with the issue of charging EVs from the standpoint of the emission of polluting gases. The objective function used in this reference is to minimize the emission of $CO_2$, and the costs related to charging EVs.

In [14], a model for the synchronous charging of EVs is presented, in which the definition of two coefficients called the capacity margin coefficient and the charging priority coefficient are used. The model used in [15], is a multi-layer composite programming model that is used for the smart charging of EVs in the DN. The objective function of the above model minimizes the total charging costs and the ratio of the peak load to the average load. In [16], the problem of charging EVs in the presence of distributed generation resources is done. The model used in this reference is a multi-objective optimization problem whose objective function includes operating costs, pollutant reduction, and load changes. The reference [17] presents a smart charging method in which the amount of energy received from the upstream network is minimized and the amount of energy produced by renewable energy sources is maximized. In [18], a multi-objective optimization problem is presented in the presence of EVs, in which the cost of energy production and the emission of polluting gases is minimized. The reference [19] provides a scenario-based optimization model to solve the problem of placing scattered production resources in the presence of renewable energy sources of electric vehicles. In [20], the types of EVs charging methods have been examined in detail.

As mentioned, the controlled charge methods are proposed against the uncontrolled charge method. In the uncontrolled charging method, it is assumed that the owner of every EV will connect his car to electricity and start charging the car as soon as he gets home. Considering that the time when people arrive at home is usually during peak hours (16:00 to 18:00), therefore using the uncontrolled charging method will put a lot of pressure on the DN. Therefore, the use of charging methods controlled or intelligent will be very effective. Because in such methods, the information related to the network is collected at every moment and sent to the central processor, and finally the charging time and also the charging rate of each EV are determined. It is necessary to implement smart methods, the existence of telecommunication, measurement, and processing infrastructures, the construction of which has very high costs.

In this article, a Semi-Smart Method (SSM) is presented in which a simple, practical, and inexpensive automation system is used to charge electric vehicles in times of network shortage. Using the proposed method, on the one hand, will reduce the negative impact of charging EVs during peak times on network parameters, and on the other hand, it will make the owner of an EV pay the cost of charging his car based on the off-peak tariff. To accurately model, the DN and arrive at the real answers, the network is considered as an unbalanced 4-wire network and the effect of the presence of the neutral wire in the charging of EVs is investigated. This is a topic that has been addressed in few articles. In better words, in the research carried out in this field, usually, the DN is either considered balanced or if it is unbalanced, the cable is not modeled. The arrival and departure time of the car, the initial charge, and the capacity of the car's battery are taken into account, and the Monte Carlo method and appropriate probability functions are used to model them. To check the efficiency of the presented method, a sample DN is used and the proposed SSM charging is compared with the controlled charging method and some other charging strategies. In each of the DN, different parameters of the network including distribution network losses, voltage profile, maximum voltage drop, and also the amount of neutral wire's voltage in different buses of the network are calculated. The obtained results confirm the effectiveness of the proposed method. Therefore, the innovations of this article are:

1. Providing a SSM for charging EVs that has the following advantages:
   - Inexpensive, without the need for various measurement, communication, and processing equipment, and infrastructures.
   - Prevention of the adverse effect of uncontrolled charging of EVs during peak load times on network parameters.
   - Minimizing the cost of charging electric cars.
2. Considering an unbalanced 3-phase 4-wire DN to observe the effect of the presence of the neutral wire and arrive at real answers.

## II. DISTRIBUTION NETWORK MODELING

DNs usually have a ring structure that is used radially. The current consumption in the DN is single-phase and 3-phase, and due to the different behavior of existing subscribers, the electric load of the network is generally unbalanced. As long as the network load is balanced, no current will pass through the loom wire. But in the unbalanced state, the current of the neutral wire is opposite to zero, and this issue causes the difference of the neutral voltage in different buses of the DN. Since a load of DN is unbalanced, it should be considered as a quadrilateral for accurate modeling of this network and to reach real answers. Figure 1 shows the equivalent circuit related to two consecutive buses in a 4-wire radial DN. According to this figure, the equations related to the voltage drop in the phase lines (ph) and neutral (n) of two consecutive buses i and j at time t can be considered in the following form.

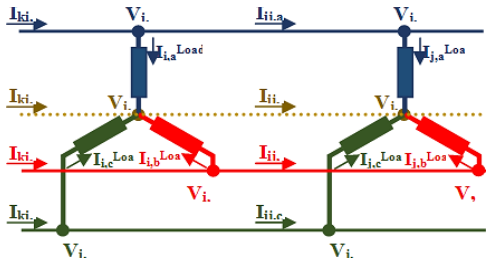

Fig. 1. The equivalent circuit related to two consecutive buses in a 4-wire radial DN LV.

$$\begin{cases} V_{i,ph,t} - V_{j,ph,t} = Z_{ij,ph} I_{ij,ph,t} \\ \forall_{i,j} \ \forall \in B \quad ph \in PH \quad t \in T \end{cases} \quad (1)$$

$$\begin{cases} V_{i,n,t} - V_{j,n,t} = Z_{ij,n} I_{ij,n,t} \\ \forall i,j \in B \quad t \in T \end{cases} \quad (2)$$

Also, the equations of the phase current and the voltage related to the bus in time will be as follows

$$\begin{cases} \sum_{\substack{k=1 \\ k<i}}^{N} I_{ki,ph,t} - I_{i,ph,t}^{Load} = \sum_{\substack{j=1 \\ j>i}}^{N} I_{ij,ph,t} \\ \forall i,j,k \in B \quad ij, ki \in L \quad ph \in PH \quad t \in T \end{cases} \quad (3)$$

$$\begin{cases} \sum_{\substack{k=1 \\ k<i}}^{N} I_{ki,n,t} + \sum_{ph} I_{i,ph,t}^{Load} = \sum_{\substack{j=1 \\ j>i}}^{N} I_{ij,n,t} \\ \forall i,j,k \in B \quad ij, ki \in L \quad ph \in PH \quad t \in T \end{cases} \quad (4)$$

$$\begin{cases} \sum_{\substack{k=1 \\ k<i}}^{N} I_{ij,ph,t} + I_{ij,n,t} = 0 \\ \forall i,j,k \in B \quad ij \in L \quad ph \in PH \quad t \in T \end{cases} \quad (5)$$

The amounts of the active power, and the reactive power in each node are determined based on the voltage and current of the same node using the following relations:

$$\begin{cases} P_{i,ph,t}^{Load} + jQ_{i,ph,t}^{Load} = (V_{i,ph\ n,t})(I_{i,ph,t}^{Load})^* \\ \forall i,j \in B \quad ph \in PH \quad t \in T \end{cases} \quad (6)$$

$$\begin{cases} V_{i,ph\ n,t} = V_{i,ph,t} - V_{i,n,t} \\ \forall i \in B \quad ph \in PH \quad t \in T \end{cases} \quad (7)$$

One of the conventional methods for solving the equations of load distribution in 4-wire radial DNs is to use the backward-forward method.

## III. MODELING OF ELECTRICAL LOADS OF THE DISTRIBUTION NETWORK

In this article, electrical loads have been divided into two categories: ordinary household loads and loads related to EVs, and the modeling of each of them will be discussed in the continuation of this section.

### A. Modeling of ordinary household loads

In this article, household loads are of single-phase type and have known active and reactive power. In this way, we will face an unbalanced distribution network in which the available loads have uncertainty. To determine the load of each of the subscribers, a basic load curve similar to Figure 2 is considered. For this purpose, the value of the load curve in each time period has been considered as the average value of the normal distribution function and the standard deviation equal to 20%.

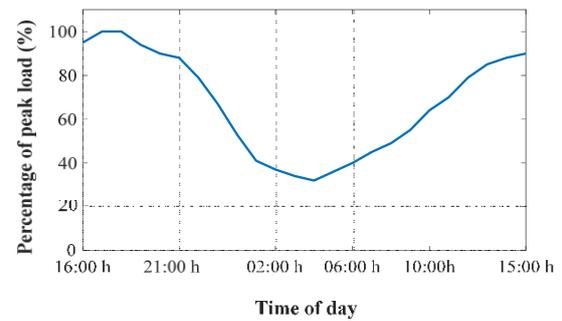

Fig. 2. Base load curve for ordinary household loads

## B. Modeling of EVs

Another type of loads considered in this article are EVs, which are considered among controllable electric loads. Electric cars, like ordinary household goods, have various uncertainties, including: battery capacity, time of entering the house, amount of initial charge and time of leaving the house. In this article, for the modeling of the above parameters, the same method is used to determine the curve of the household load of the subscribers. with the difference that instead of the normal distribution function, appropriate probability distribution functions have been used for each of the above parameters based on the information of the available articles in this field.

Another important point in the modeling of EVs is determining their power consumption when charging. Usually, two types of slow and fast chargers are used to charge EVs. In the slow charging method, which is usually done in homes, the current drawn by the EV is equal to 16 A or 32 A under a voltage of 220 V, which is equivalent to power. 3.5 kW and 7 kW. In fast charging, the consumption power of the EV increases up to 50 kW or higher, which is usually used in special parking lots for charging EVs. In addition, in order to reach the optimal values of various parameters of the smart DN, the charging power rate can also be controlled. But in this article, it is assumed that the network lacks special communication and control infrastructures, and for this reason, its value is 3.5 kW.

Considering that EVs use only active power, $P_{i,ph,t}^{Load}$ and $Q_{i,ph,t}^{Load}$ in (6) can be considered as the following relations:

$$\begin{cases} P_{i,ph,t}^{Load} = P_{i,ph,t}^{Conv} + P_{i,ph,t}^{PHEV} \\ \forall i \in B \quad ph \in PH \quad t \in T \end{cases} \quad (8)$$

$$\begin{cases} Q_{i,ph,t}^{Load} = Q_{i,ph,t}^{Conv} \\ \forall i \in B \quad ph \in PH \quad t \in T \end{cases} \quad (9)$$

When each car arrives at home and knowing the initial charge of the car battery, battery capacity, and charging power, the time needed to fully charge the EV can be obtained from the relationship below.

$$\begin{cases} TCT_{i,ph} = \dfrac{BC_{i,ph} \times (0.95 - ISOC_{i,ph})}{ChP} \\ \quad , \forall i \in B \quad ph \in PH \end{cases} \quad (10)$$

It should be mentioned that usually, to prevent battery damage and reduce its useful life, the battery is charged up to 95% of its capacity. The main point after determining the time required to charge EVs, is determining the time to charge the car. In the next section, a SSM for EV charging will be presented.

## IV. SEMI-SMART CHARGING OF ELECTRIC VEHICLES

As mentioned in the previous sections, two methods are usually used to charge EVs, which are: (1) controlled or smart method; 2) uncontrolled or random method. In the uncontrolled method, every EV starts charging as soon as it enters the house and connects to the network. In this case, there is no need for any telecommunications infrastructure to control the charging of cars, and considering the random behavior of car owners, it is non-deterministic in nature. But in controlled mode, the goal is: "Determining the optimal charging time and power rate for charging EVs in the network." Specifically, in the smart method, the existence of communication platforms and smart measurement equipment and a central processing and control system is unavoidable. The method of semi-smart charging of EVs is presented, which does not require network smart and the use of expensive equipment on the one hand, and on the other hand, avoids the negative effects of uncontrolled charging of EVs during peak load times.

As can be seen from Figure 2, the load curve of a typical household consumer can be divided into three areas: 1) low load, 2) intermediate load, and (3) high load or peak load. In order to avoid putting pressure on the DN during peak times, generally, electricity tariff is also divided into three regions according to Figure 3, which are: (1) green region (low tariff), (2) blue region (medium tariff) and (3) red region (high tariff).

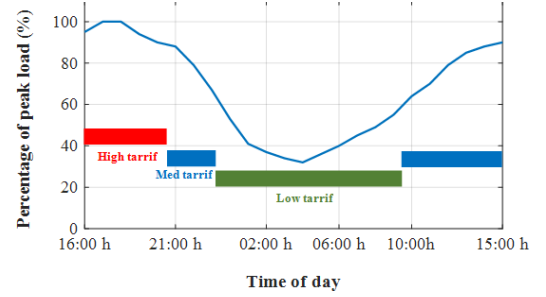

Fig. 3. A sample of energy tariff based on the load curve of normal household consumption

As it can be seen from Figure 3, charging of EVs in the low load area on the one hand will cause less pressure on the DN and as a result of improving the parameters of the DN, and on the other hand, it will have a lower charging cost for the car owners. This is when most car owners prefer to connect their car to the network as soon as they reach home so that the car is ready for the next morning with a full charge. Since the charging is not controlled, it puts a lot of pressure on the network and the use of the smart charging method is also expensive, in this article a SSM is presented in which, by using cheap equipment such as programmable key EVs are charged during times of network shortage. The working method is as follows: as soon as the car owner reaches home, he connects his car to the network through a programmable key, and the information related to the time of leaving the house, the initial charge, and it enters the battery capacity of its car as input information to the key. According to the input information and using (8), the time required to fully charge the EV's battery is calculated by the key processor and the EV charging time is determined in such a way that the end of it is the moment the car owner leaves the house in the morning. Be later the charging start time is calculated based on the time needed to charge the car and also the charging power rate of the car or the same 3.5 kW. In other words, the amount $P_{i,ph,t}^{Load}$ for all EVs is determined locally by programmable keys installed in the parking lots of homes and based on the input information of the car owner, using the below equations:

$$\begin{cases} P_{i,ph,t}^{Load} = ChP, \quad Dep_{i,ph} - TCT_{i,ph} \le t \le Dep_{i,ph} \\ \quad 0, \qquad \qquad Otherwide \end{cases} \quad (11)$$

The reason for this choice is that usually the time to leave the house is around 7 a.m., which is comparable to the low network times. Therefore, in addition to the electricity distribution company, charging the car in the garages located in the low tariff area will also benefit the car owner; because the price of electric power is calculated based on the low tariff. Therefore, using the proposed semi-smart charging method has the following advantages:

## V. SIMULATION

To check and evaluate the efficiency of the proposed method, a 380V three-phase 19-bus feeder was used according to figure 4, and the information related to the resistance and inductance of the phase lines is given in Table I. The specifications of the neutral conductor are also included. It is assumed that in each bus, three home consumers are connected to each of the phases, and therefore the total number of subscribers in the network is equal to 57. The studied period is a day and night which is divided into 15-minute intervals. The load curve of the household sharers has been determined according to the method presented in section (3-A), and the peak load and common power factor have been taken into account as 2 kW and 0.91 lead-phase, respectively. In order to assign EVs to network subscribers, the working method is in the order that the penetration coefficient of EVs is considered equal to 60%. In other words, 60% of household subscribers, that is, 34 out of 57 subscribers located in different phases, are the owners of EVs, which were randomly selected. To model the uncertainty in the parameters of EVs, the same method presented in section (3-A) is used for modeling household loads, with the difference that instead of the normal function, the probability distribution functions shown in Table II are used. Table III shows the locations of randomly selected EVs, and their various parameters based on the above explanations. As it can be seen from Table III, for example, in bus 1, only the users of phases a and b have electric cars, whose battery capacity is equal to 26 and 19 kwh, respectively.

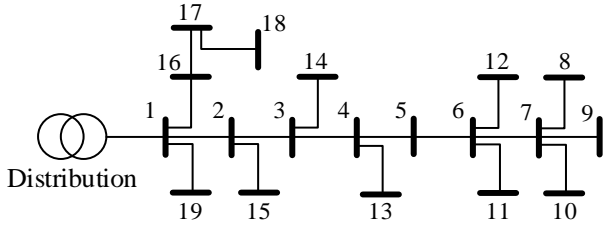

Fig. 4. LV feeder for distribution of 19 buses of the DN

TABLE I. SPECIFICATIONS OF FEEDER LINES WITH 19 BUSES

| Bus | | Resistance of Line (Ω) | Inductance of Line (Ω) |
|---|---|---|---|
| From | To | | |
| 1 | 2 | 0.0415 | 0.0145 |
| 2 | 3 | 0.0424 | 0.0189 |
| 3 | 4 | 0.0444 | 0.0198 |
| 4 | 5 | 0.0369 | 0.0165 |
| 5 | 6 | 0.0520 | 0.0232 |
| 6 | 7 | 0.0524 | 0.0234 |
| 7 | 8 | 0.0005 | 0.0002 |
| 7 | 9 | 0.2002 | 0.0199 |
| 7 | 10 | 1.7340 | 0.1729 |
| 6 | 11 | 0.2607 | 0.0260 |
| 6 | 12 | 1.3605 | 0.1357 |
| 4 | 13 | 0.1400 | 0.0140 |
| 3 | 14 | 0.7763 | 0.0774 |
| 2 | 15 | 0.5977 | 0.0596 |
| 1 | 16 | 0.1423 | 0.0496 |
| 16 | 17 | 0.0837 | 0.0292 |
| 17 | 18 | 0.3123 | 0.0311 |
| 1 | 19 | 0.0163 | 0.0062 |
| Distribution's transformer | | | 0.0654 |

TABLE II. THE PARAMETERS OF EV

| Parameter | Probability distribution function | average value | standard deviation | Minimum amount | maximum value |
|---|---|---|---|---|---|
| Capacity (kWh) | monotone | 18 | 6.93 | 6 | 30 |
| Arrival time (h) | Normal is disconnected | 19 | 2 | 16 | 1 |
| Departure time (h) | Normal is disconnected | 7 | 2 | 5 | 12 |
| Initial charge (%) | Normal is disconnected | 75 | 25 | 25 | 95 |

TABLE III. LOCATION OF EVs AND VARIOUS PARAMETERS RELATED TO THEM

| Bus.Phase | Capacitor (kWh) | arrival time (h) | departure time (h) | Initial charge (%) |
|---|---|---|---|---|
| 1.a | 26 | 17:00 | 05:30 | 65 |
| 1.b | 19 | 23:15 | 05:00 | 65 |
| 2.a | 30 | 18:00 | 07:45 | 17 |
| 2.b | 8 | 16:30 | 09:30 | 14 |
| 3.b | 17 | 19:00 | 06:15 | 42 |
| 3.c | 9 | 18:45 | 05:45 | 70 |
| 4.a | 29 | 21:00 | 05:30 | 64 |
| 4.c | 25 | 20:00 | 11:45 | 60 |
| 5.a | 26 | 18:15 | 08:00 | 31 |
| 5.c | 8 | 20:45 | 09:15 | 49 |
| 6.c | 25 | 21:00 | 08:45 | 65 |
| 7.a | 16 | 19:30 | 05:45 | 48 |
| 7.b | 28 | 16:00 | 08:30 | 5 |
| 7.c | 10 | 18:45 | 10:15 | 25 |
| 8.b | 9 | 18:00 | 09:00 | 49 |
| 8.c | 9 | 20:30 | 05:45 | 22 |
| 9.a | 27 | 18:00 | 08:45 | 34 |
| 9.b | 20 | 17:00 | 11:30 | 56 |
| 9.c | 19 | 18:15 | 08:00 | 25 |
| 10.b | 26 | 20:30 | 07:45 | 34 |
| 11.a | 14 | 20:45 | 08:00 | 23 |
| 12.a | 8 | 19:30 | 06:30 | 32 |
| 12.c | 9 | 17:15 | 06:15 | 60 |
| 13.a | 10 | 16:00 | 09:30 | 74 |
| 13.c | 16 | 19:15 | 07:00 | 41 |
| 14.c | 29 | 23:00 | 08:00 | 53 |
| 15.a | 18 | 19:00 | 07:45 | 64 |
| 15.b | 18 | 20:45 | 05:30 | 60 |
| 16.c | 9 | 20:15 | 06:00 | 35 |
| 17.a | 25 | 17:30 | 07:15 | 54 |
| 17.b | 15 | 16:45 | 05:45 | 42 |
| 18.a | 16 | 16:30 | 06:30 | 65 |
| 18.b | 8 | 19:00 | 05:45 | 76 |
| 19.c | 20 | 21:45 | 09:00 | 56 |

To observe the effect of the presence of EVs on the network parameters, first the network is simulated without the presence of EVs, which is called the base case. Also, the results related to the presented SSM for EV charging will be compared with three other methods. Therefore, simulations are performed in five different modes, which are:

A. First case: DN, without the presence of an EVs

B. Second case: DN with the presence of EVs, and the charging of vehicles in an uncontrolled method.

C. Third mode: DN with the presence of EVs, and charging cars using a timer key. In this case, it is assumed that all cars start charging at 24:00.

D. The fourth mode: DN with the presence of EVs and car charging using a timer key and DN zoning. In this case, it is assumed that the DN is divided into 3 regions and the cars of each region start charging at 23:00 with an interval of one hour. The way of division into regions and the start time of charging in each region is shown in Table IV.

E. Fifth mode: DN with the presence of EVs and vehicle charging using a programmable key.

In all three cases, the backward-forward method has been used to play the load in each time period. The phase voltage of bus 1 is considered to be equal to 220 volts and the neutral voltage in bus 1 is equal to zero.

TABLE IV. HOW TO DIVIDE THE REGIONS AND ALSO THE CHARGING START TIME OF EACH REGION IN THE FOURTH STATE

| Area's number | number of buses available in each area | Charging start time |
|---|---|---|
| 1 | 1,2,15,16,17,18,19 | 23:30 |
| 2 | 3,4,5,6,13,14 | 24:00 |
| 3 | 7,8,9,10,11,12 | 01:00 |

The first parameter that is used to compare the SSM with uncontrolled charging is the voltage profile along the feeder. In figures 5 to 7, the changes of the voltage of different phases in relation to the zero for the worst bus in the five described cases are shown. It is necessary to mention that the worst bus is the bus where the highest amount of voltage drop occurs. The voltage values in the worst bus for different conditions are shown in Table V.

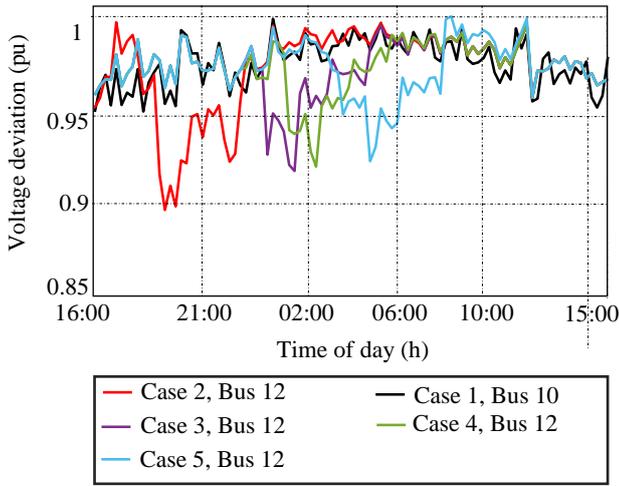

Fig. 5. Voltage changes of phase (A) compared to zero in the worst bus

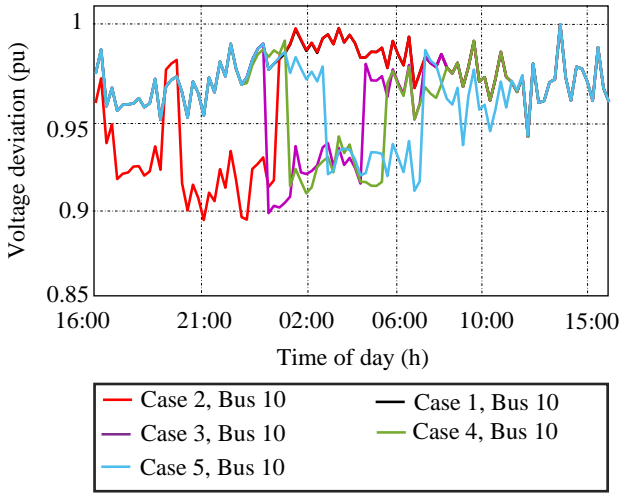

Fig. 6. Voltage changes of phase (B) compared to zero in the worst bus

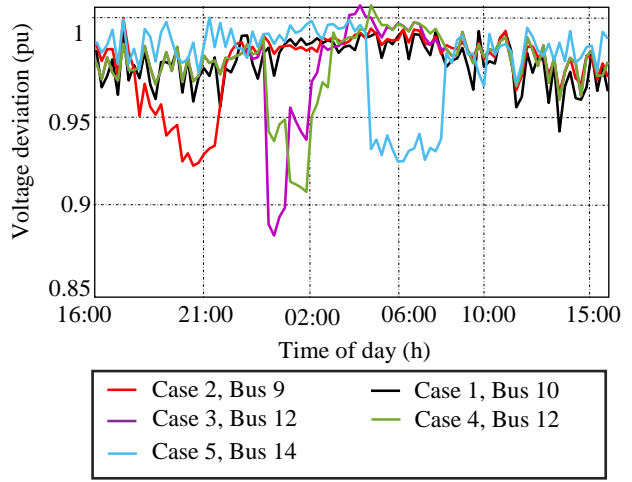

Fig. 7. Voltage changes of phase (C) compared to zero in the worst bus

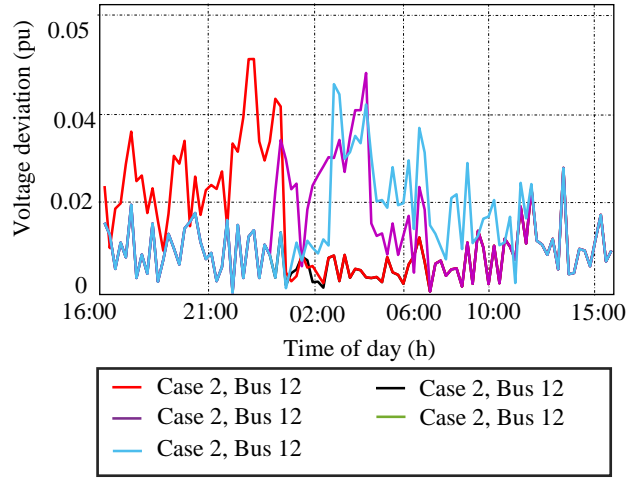

Fig. 8. Voltage changes of Neutral wire (N) in the worst bus

TABLE V. VOLTAGE VALUES IN THE WORST CASE FOR DIFFERENT CONDITIONS

| State | The value of the voltage in the worst bus (pu) | | | The lowest voltage value (pu) |
|---|---|---|---|---|
| | $V_{an}$ | $V_{bn}$ | $V_{cn}$ | |
| A | 0.9489 | 0.9409 | 0.9389 | 0.9389 |
| B | 0.8962 | 0.8963 | 0.9205 | 0.8962 |
| C | 0.9172 | 0.9000 | 0.8829 | 0.8829 |
| D | 0.9195 | 0.9106 | 0.9064 | 0.9064 |
| E | 0.9225 | 0.9121 | 0.9229 | 0.9121 |

As it is clear from the results, the highest amount of voltage drop is related to the case where the cars start charging by using the timer switch and at a specific time, the same 24:00. This issue can be solved by dividing the network into three zones and allocating separate charging times to each zone, and it is observed that by using this method, the amount of voltage drop is improved compared to the uncontrolled charging method. It is also observed that in order to reach a better answer, a programmable key can be used and the amount of voltage drop can be brought to an acceptable value. The zero-voltage value is another parameter whose changes in the worst bus have been compared for different conditions and the results are shown in Figure 8. It is necessary to mention that the worst bus is meant here, the bus where the highest amount of zero voltage occurs. As expected, due to the load imbalance of the network, the current passing through the neutral wire is not zero, and as a result, the voltage of the neutral wire will also be zero in the different buses of the opposite network. Uncontrolled entry of EVs into the network causes the network to become more unbalanced and the amount of zero voltage to increase. However, it is observed that the amount of zero voltage is less in the case of using a programmable key than in other cases. To clarify, it should be noted that in the uncontrolled method, the charging of cars is uneven and happens randomly during the peak times of the network. In the methods related to the use of a timer key, there is also a kind of synchronization between the entry of large unbalanced loads into the network, which can increase the level of imbalance of the basic network. However, in the presented method, the charging of the cars is done asynchronously in times of network shortage, and this results in obtaining better results in comparison with other charging methods. Network losses are among other parameters that are very important in the distribution network. Figure 9 shows the amount of total wasted energy related to the first to fifth modes during day and night. As can be seen, the amount of wasted energy in the absence of EVs is equal to 192.803 kwh, which increases to 287.249 kwh when EVs enter the grid and their controlled charging. By using SSM and charging EVs during off-peak hours, the amount of network losses is reduced, so that with the use of the timer switch, the value is 271.949 kwh, and with the use of the timer switch and zoning of the network, it decreases to 265.830 kwh. This means that by using simple equipment and without network intelligence, the amount of network losses will be improved by 5.6% and 8% respectively by using the above methods. Finally, by using the programmable key, the amount of network losses reaches 256.240 kwh, which means a 12.1% reduction in network losses. It is observed that the semi-smart charging method, using a programmable key, improves the network parameters in the presence of EVs compared to the uncontrolled charging method, and makes the car charging costs for the owners as low as possible. It is necessary to mention that by generalizing the presented model, the ability to inject power from the Vehicle to the Grid (V2G) can also be added to it. This is a topic that is being researched.

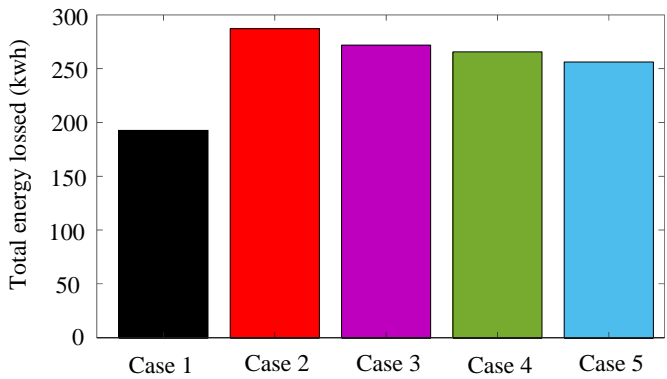

Fig. 9. The amount of wasted energy during the day and night

## VI. ONCLOTION

In this article, a SSM for charging EVs in the DN has been presented. Compared to smart methods, this method has a very small cost, and compared to the uncontrolled charging method, it improves the parameters of the DN in the presence of EVs. In this method, it is assumed that the charging rate of EVs is fixed and the goal is to determine the charging interval of the car. To do this, by using a programmable key, the EV charging time is determined in such a way that the end of the time is the car's departure time in the morning of the next day. Therefore, it is necessary for the car owner to enter his car information as an input to the programmable key. By applying the presented method on the 19-bus DN and comparing the results obtained from it with different uncontrolled methods, its efficiency has been evaluated. In any case, the main network parameters including losses, voltage drop and zero voltage have been calculated. The obtained results show that the use of the proposed method leads to an improvement of 12.1% in the amount of network losses and 3.3% in the amount of network voltage drop. The important point is that using the proposed method will minimize the cost of charging EVs for their owners, because in the presented method, EV charging is done in times of network shortage, in which the cost of the power of many consumers is high.

*List of symptoms*

| | |
|---|---|
| $i, j, k$ | Indications related to network buses |
| $ph$ | Indication related to phase |
| $n$ | Indication of the neutral wire |
| $t$ | Indication related to time period |
| $B$ | Set of all network buses |
| $L$ | Set of all network lines |
| $PH$ | Set of all phases of the network {a, b, c} |
| $T$ | Set of all time intervals |
| $BC_{i,ph}$ | Capacity of the battery of the EV placed in the ph and i-th bus |
| $ChP$ | Car charging power rate (3.5 kW) |
| $Dep_{i,ph}$ | Departure time of the EV located in ph and i-th bus in the next morning |
| $ISOC_{i,ph}$ | Value of the initial charge of the EV placed in the ph and i-th bus |
| $P_{i,ph,t}^{Conv.}$ | Active power of a normal household load is located in the ph and i-th bus at time t |
| $Q_{i,ph,t}^{Conv.}$ | Reactive power of a normal household load is located in the ph and i at time t |
| $Z_{ij,ph}$ | Impedance of the ph between the i-th bus and j-th bus |
| $Z_{ij,n}$ | Impedance of the null conductor between the i-th bus and j-th bus |
| $I_{ij,ph,t}$ | Current of the ph between the i-th bus and j-th bus at time t |
| $I_{ij,n,t}$ | Current of the neutral conductor between the i-th bus and j-th bus at time t |
| $I_{i,ph,t}^{Load}$ | Electric charge current between the ph and zero bus at time t |
| $P_{i,ph,t}^{Load}$ | Active power of the electric charge is located between the ph and zero bus at time t |
| $Q_{i,ph,t}^{Load}$ | Reactive power of the electric charge is located between the ph and zero bus in the at time t |
| $TCT_{i,ph}$ | The time required to charge the car |
| $V_{i,ph,t}$ | Phase voltage ph of the i-th bus at time t |
| $V_{i,n,t}$ | Neutral conductor's voltage of the i-th bus at time t |
| $V_{i,ph\,n,t}$ | The difference in the potential of the ph of i-th bus compared to neutral conductor |